\documentclass[floatfix,prl,twocolumn,letterpaper,lengthcheck,superscriptaddress,showpacs,amssymb,amsmath,amsfonts,aps,altaffilletter,nofootinbib,nopreprintnumbers,showpacs,longbibliography]{revtex4-2}

\usepackage{times}
\usepackage{color}
\usepackage{hyperref}
\usepackage{multirow}
\usepackage{graphicx}
\usepackage{placeins}
\usepackage{tabularx}
\usepackage{url}
\usepackage{xcolor}

\begin{document}

\title{A multimode quasi-normal spectrum from a perturbed black hole}

\author{Collin D. Capano}
\affiliation{Max-Planck-Institut f{\"u}r Gravitationsphysik
  (Albert-Einstein-Institut),Callinstra{\ss}e 38, 30167 Hannover,
  Germany}
\affiliation{Leibniz Universit{\"a}t Hannover, 30167 Hannover, Germany}
\affiliation{Department of Physics, University of
  Massachusetts,Dartmouth, MA 02747, USA}

\author{Miriam Cabero}
\affiliation{Department of Physics and Astronomy, The University of
  British Columbia,Vancouver, BC V6T 1Z4, Canada}

\author{Julian Westerweck}
\affiliation{Max-Planck-Institut f{\"u}r Gravitationsphysik
  (Albert-Einstein-Institut),Callinstra{\ss}e 38, 30167 Hannover,
  Germany}
\affiliation{Leibniz Universit{\"a}t Hannover, 30167 Hannover, Germany}

\author{Jahed Abedi}
\affiliation{Max-Planck-Institut f{\"u}r Gravitationsphysik
  (Albert-Einstein-Institut),Callinstra{\ss}e 38, 30167 Hannover,
  Germany}
\affiliation{Leibniz Universit{\"a}t Hannover, 30167 Hannover, Germany}
\affiliation{Department of Mathematics and Physics, University of
  Stavanger,NO-4036 Stavanger, Norway}

\author{Shilpa Kastha}
\affiliation{
  Niels Bohr International Academy, Niels Bohr Institute,
  Blegdamsvej 17, 2100 Copenhagen, Denmark
}
\affiliation{Max-Planck-Institut f{\"u}r Gravitationsphysik
  (Albert-Einstein-Institut),Callinstra{\ss}e 38, 30167 Hannover, Germany}
\affiliation{Leibniz Universit{\"a}t Hannover, 30167 Hannover, Germany}

\author{Alexander H. Nitz}
\affiliation{Max-Planck-Institut f{\"u}r Gravitationsphysik
  (Albert-Einstein-Institut),Callinstra{\ss}e 38, 30167 Hannover,
  Germany}
\affiliation{Leibniz Universit{\"a}t Hannover, 30167 Hannover, Germany}

\author{Yi-Fan Wang}
\affiliation{Max-Planck-Institut f{\"u}r Gravitationsphysik
  (Albert-Einstein-Institut),Callinstra{\ss}e 38, 30167 Hannover,
  Germany}
\affiliation{Leibniz Universit{\"a}t Hannover, 30167 Hannover, Germany}

\author{Alex B. Nielsen}
\affiliation{Department of Mathematics and Physics, University of
  Stavanger,NO-4036 Stavanger, Norway}

\author{Badri Krishnan}
\affiliation{Max-Planck-Institut f{\"u}r Gravitationsphysik
  (Albert-Einstein-Institut),Callinstra{\ss}e 38, 30167 Hannover,
  Germany}
\affiliation{Leibniz Universit{\"a}t Hannover, 30167 Hannover, Germany}
\affiliation{Institute for Mathematics, Astrophysics and Particle
  Physics, Radboud University,Heyendaalseweg 135, 6525 AJ Nijmegen,
  The Netherlands}

\def\l{\ell}
\def\lmn{\ell mn}
\def\mdot{\mathrm{M}_{\odot}}

\def\BFthreethree{{\ensuremath{56\pm1}}}
\def\BFovertone{{\ensuremath{8.2^{+0.7}_{-0.7}\times 10^{4}}}}
\def\BFdetection{{56\pm1}}
\def\estMf{\ensuremath{330_{-40}^{+30}}}
\def\estSf{\ensuremath{0.86_{-0.11}^{+0.06}}}
\def\dfThreeThreeMain{\ensuremath{-0.01^{+0.08}_{-0.09}}}
\def\dTauThreeThreeMain{\ensuremath{0.6^{+1.9}_{-1.2}}}
\def\dfOvertone{\ensuremath{-0.10^{+0.34}_{-0.05}}}
\def\dTauOvertone{\ensuremath{0.5^{+0.3}_{-0.7}}}

\def\agnosticSNRA{{\ensuremath{12.2}}}
\def\agnosticSNRB{{\ensuremath{4.1}}}
\def\qthreethree{{\ensuremath{0.4^{+0.2}_{-0.3}}}}
\def\agnosticFreqA{{\ensuremath{63^{+2}_{-2}\,\rm{Hz}}}}
\def\agnosticTauA{{\ensuremath{26^{+8}_{-6}\,\rm{ms}}}}
\def\agnosticFreqB{{\ensuremath{98^{+89}_{-7}\,\rm{Hz}}}}
\def\agnosticTauB{{\ensuremath{40^{+50}_{-30}\,\rm{ms}}}}

\def\ampThreeThree{\ensuremath{0.2^{+0.2}_{-0.1}}}
\def\threethreeSNR{\ensuremath{12.6}}

\begin{abstract}

  When two black holes merge, the late stage of gravitational wave
  emission is a superposition of exponentially damped
  sinusoids. According to the black hole no-hair theorem, this
  ringdown spectrum depends only on the mass and angular momentum of
  the final black hole. An observation of more than one ringdown mode
  can test this fundamental prediction of general relativity. Here we
  provide strong observational evidence for a multimode black hole
  ringdown spectrum using the gravitational wave event GW190521, with
  a maximum Bayes factor of $\BFdetection{}$
  ($1\sigma$ uncertainty) preferring two fundamental modes over
  one. The dominant mode is the $\ell=m=2$ harmonic, and the
  sub-dominant mode corresponds to the $\ell=m=3$ harmonic. The amplitude of
  this mode relative to the dominant harmonic is estimated to be
  $A_{330}/A_{220} = \ampThreeThree{}$. We
  estimate the redshifted mass and dimensionless spin of the final
  black hole as $\estMf~\mdot$ and $\estSf$, respectively.
  We find that the final black hole is consistent with the no hair theorem and
  constrain the fractional deviation from general relativity of the
  sub-dominant mode's frequency to be \dfThreeThreeMain{}.
   
\end{abstract}

\maketitle

\section{Introduction}

A perturbed black hole approaches equilibrium by emitting a spectrum of damped sinusoidal gravitational-wave signals~\cite{Vishveshwara:1970cc,Chandrasekhar:1975zza,TheLIGOScientific:2016src}. Unlike other astrophysical objects, the ringdown spectrum of a black hole is remarkably simple. General relativity predicts that the frequencies and damping times of the entire spectrum of damped sinusoids, or ``quasi-normal modes'', are fully determined by just two numbers: the black hole mass $M$ and angular momentum $J$, as described by the Kerr solution~\cite{Kerr:1963ud}.  This prediction, a consequence of the black hole ``no-hair theorem", does not hold in many alternate theories~\cite{Cardoso:2019rvt}. If astrophysical black holes are observed to violate this property, it indicates new physics beyond standard general relativity.

In order to observationally test this prediction using binary black hole mergers, an important observational challenge must be met: at least two ringdown modes must be observed~\cite{Dreyer:2003bv}. The higher the binary mass ratio asymmetry, the more likely it is that sub-dominant ringdown modes are observable.  However, more asymmetric binary systems are less likely to be formed, and also lead to weaker signals. Population studies suggested that such multimode ringdown modes were unlikely to be observed until the next generation of gravitational-wave observatories~\cite{Berti:2016lat,Cabero:2019zyt}, since black-hole population models did not anticipate observations of massive, asymmetric binaries.

Here we confound this expectation with the gravitational-wave event GW190521, detected by the two LIGO detectors and Virgo at 03:02:29 UTC on May 21st, 2019~\cite{Abbott:2020tfl,Abbott:2020mjq}. This is the heaviest confidently detected black-hole merger event observed to date~\cite{Abbott:2020niy,Nitz:2021uxj}. The signal is consistent with the merger of two high mass black holes which merge at a low frequency relative to the detector sensitivity band. As such, it has a barely observable inspiral; the signal is dominated by the merger and ringdown phase.

GW190521 was initially reported as the merger of two comparable mass black holes~\cite{Abbott:2020tfl,Abbott:2020mjq}. If the spins are aligned, one would not expect to detect sub-dominant ringdown modes. However, there was significant posterior support for precession in GW190521. Furthermore, subsequent re-analysis of the data found that the progenitor masses could have been unequal~\cite{Nitz:2020mga, Estelles:2021jnz}. Both of these scenarios suggest the possibility of detectable sub-dominant modes~\cite{JimenezForteza:2020cve,Siegel:2023lxl}. Here we find strong evidence for multimode damped sinusoids in the ringdown phase of the gravitational wave event GW190521.

\section{Multimode agnostic search}
%***************************************************
\label{sec1:agnostic}

\begin{figure*}
\includegraphics[width=\textwidth]{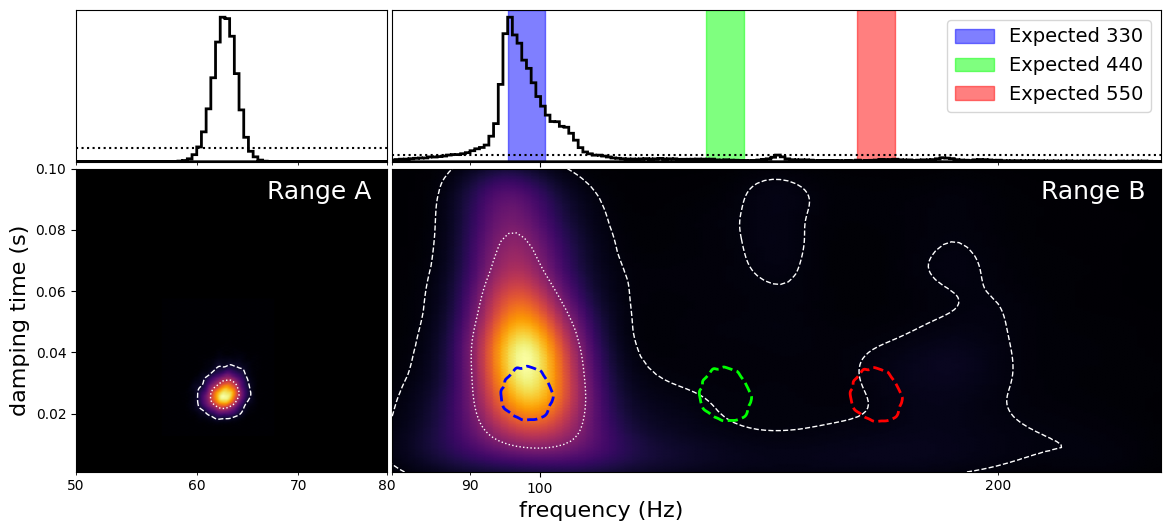}
\caption{Marginal posterior probability distributions on frequency and damping time from an agnostic quasi-normal mode analysis of GW190521 at 6ms after $t_{\rm ref}$. A single mode is searched for in each of the shown frequency ranges, range A (50 to 80Hz) and range B (80 to 256Hz). Top panels show the marginal posterior on the mode frequencies, with priors indicated by dotted lines. White dotted (dashed) contours in the bottom panels show the 50th (90th) credible regions. Assuming the dominant mode in range A corresponds to the (220) mode of a Kerr black hole, we estimate what the frequency and damping times would be of the (330), (440), and (550) modes (blue, green, and red regions, respectively). The mode in range B is clearly consistent with the expected frequency and damping time of the (330) mode. Here we do not see the (440) and (550) modes, indicating they are weaker than the (330) mode. This is consistent with an asymmetric binary black hole merger.} 
\label{fig:agnostic6ms}
\end{figure*}

A quasi-normal mode description of the gravitational wave from a binary black hole is not expected to be valid until after the binary has merged to form a perturbed black hole. On the flip side, the damping time of an $O(100\,\mathrm{M}_\odot)$ black hole is $O(10\,\mathrm{ms})$, leaving a window of only a few tens of milliseconds after merger in which the ringdown is detectable above noise. Accurate identification of the merger time is therefore crucial to extract quasi-normal modes from the data. To account for uncertainty in the merger time of GW190521 due to modelling systematics, we perform a series of analyses in short time increments starting at a geocentric GPS reference time $t_{\rm ref} = 1242442967.445$. This time is taken from the maximum likelihood merger time obtained via the analysis in Nitz \& Capano~\cite{Nitz:2020mga}. We also fix the sky location to the maximum likelihood values from the same analysis.

The ringdown spectrum of a Kerr black hole consists of an infinite set of frequencies $f_{\lmn }$ and damping times $\tau_{\lmn}$ labeled by three integers $(\ell,m,n)$. Here $\ell$ and $m$ are the usual angular harmonic numbers. The third index $n\geq 0$ denotes overtones, with $n=0$ being the fundamental mode.
The most agnostic way to search for quasi-normal modes from a perturbed black hole is to search for them individually, without assuming any relation between them. Such a search is complicated by the nature of quasi-normal modes: they are not orthogonal, meaning that modes that overlap in time must be sufficiently separated in frequency or damping time in order to be distinguishable. Simulations of binary black hole mergers have shown that the fundamental $\ell=m=2$ mode is typically significantly louder than other modes. In order to extract sub-dominant modes from noisy data in an agnostic search it is useful to separate the dominant mode in frequency from the others.

A visual inspection of the time- and frequency-domain data taken at the reference time revealed significant power in the two LIGO detectors between 60-70\,Hz (see the Supplemental Material). In order to isolate this and search for sub-dominant modes we constructed three frequency ranges: ``range A'' between $50-80\,$Hz, ``range B'' between $80-256\,$Hz, and ``range C'' between $15-50\,$Hz. We search for one quasi-normal mode in each range using Bayesian inference. We use uniform priors on the relative amplitudes of the modes in range B and C between 0 and 0.9 times the mode in range A. No other relation is assumed between the modes.

We repeat this analysis at time steps of $t_{\rm ref}$ + 0, 6, 12, 18, and $24\,$ms. As expected from the visual inspection of the data, we find a significant mode in range A at all grid points, which decreases in amplitude at later times. A clear second mode is found in range B. This mode is most visible at $t_{\rm ref} + 6\,$ms, the result of which is shown in Fig.~\ref{fig:agnostic6ms} (results at other times are shown in the Supplemental Material). The frequency of the secondary mode at this time is \agnosticFreqB{} with a damping time of \agnosticTauB{}, while the primary mode has a frequency of \agnosticFreqA{} and damping time \agnosticTauA{}. The signal-to-noise ratio (SNR) of the primary and secondary modes of the maximum likelihood waveform is \agnosticSNRA{} and \agnosticSNRB{}, respectively. Results from range C (not shown) are consistent with noise.

The dominant mode found at \agnosticFreqA{} is expected to be the quadrupolar $\ell=m=2,\,n=0$ fundamental mode. Measurement of $f_{220}$ and $\tau_{220}$ provides an estimate of the mass and angular momentum of the remnant black hole~\cite{Berti:2005ys}. This in turn predicts the entire ringdown spectrum of subdominant modes.  Figure \ref{fig:agnostic6ms} shows that the subdominant mode at \agnosticFreqB{} is consistent with the $\ell=m=3,\,n=0$ mode. This is also in agreement with expectations from numerical simulations of binary black hole mergers~\cite{Kamaretsos:2011um,Borhanian:2019kxt,JimenezForteza:2020cve}.

To quantify the agreement between the expected (330) mode and the observed mode in Range B, we multiply the observed posterior in range B (color map in Fig.~\ref{fig:agnostic6ms}) with the expected distribution using range A (indicated by the blue contour in Fig.~\ref{fig:agnostic6ms}) and integrate~\cite{Capano:2022zqm}. This yields a statistic $\zeta$ that is proportionate to the agreement between the expected mode and the observed mode. Repeating on all the grid points, we find that $\zeta$ obtains a maximum value of $1.27$ at $t_{\rm ref}+6\,$ms, consistent with our visual inspection. In Ref.~\cite{Capano:2022zqm} we perform a large simulation campaign in the data surrounding GW190521. Repeating the agnostic analysis on these simulations, we find that the probability of finding $\zeta \geq 1.27$ in noise is $\sim 0.004$.

\section{Consistency with the Kerr solution}
%**********************************************
\label{sec2:kerrtest}

\begin{figure}
\includegraphics[height=2.75in]{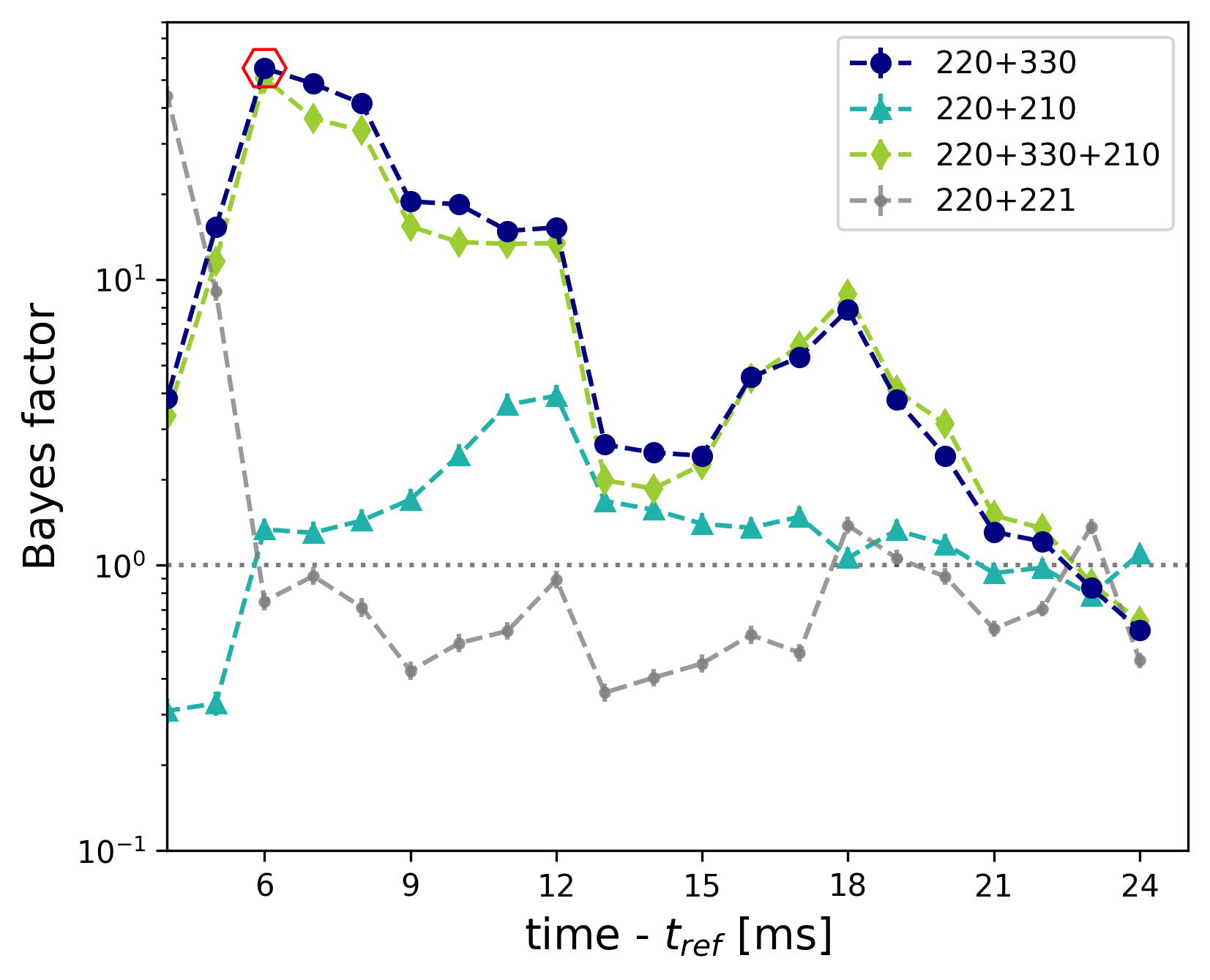} 
\caption{Bayes factor of Kerr models that include the indicated modes. The Bayes factor for the (220) + (221) model is calculated against the (220)-only model. For all other models, the Bayes factors are calculated by comparing the evidence for the model against the maximum of the (220)-only model and the (220)+(221) model. The hexagon marks the point with the largest overall Bayes factor, which is for the (220) + (330) model. Quoted mass and spin estimates are taken at this point, as well as the no-hair test described in the text.}
\label{fig:Bayes_vs_t}
\end{figure}

\begin{figure}
\includegraphics[height=2.75in]{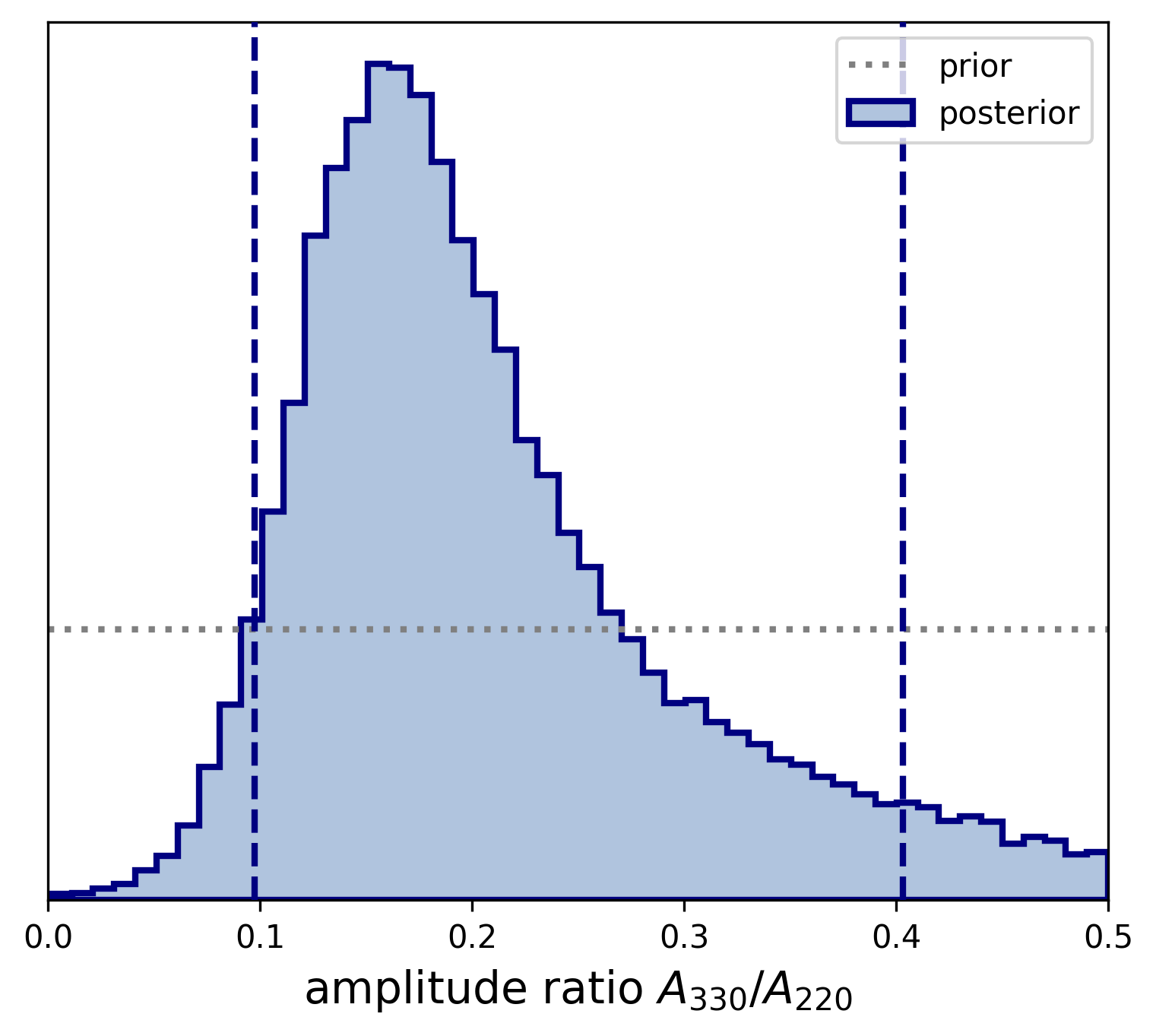}
\caption{Marginal posterior distribution of the amplitude ratio of the (330) mode relative to the (220) mode, $A_{330}/A_{220}$. The gray dotted line shows the prior, which was uniform in $[0, 0.5)$. Vertical dashed lines indicate the 90\% credible interval. The posterior distribution is obtained from the (220) + (330) model starting at $t_{\rm ref}+6\,$ms (red hexagon in Fig.~\ref{fig:Bayes_vs_t}), which is the most favored model in the Kerr analysis.}
\label{fig:amp_posterior}
\end{figure}

The search for damped sinusoids in the previous section assumed no particular relation between different modes, with a corresponding large prior parameter volume. In this section, we assume that the frequency and damping times of the damped sinusoids are related as in the ringdown of a Kerr black hole. This reduces the prior parameter volume and focuses in on particular modes. The amplitudes and phases of the modes are left as free parameters, since they depend on the specific initial state of the remnant black hole immediately after the merger.

For this analysis, we model the ringdown signal based on the final Kerr black hole mass, $M_f$, and dimensionless spin, $\chi_f = J_f/M_f^2$. We expect only a subset of the entire spectrum of quasi-normal modes to be visible above noise. Including all possible modes in our signal model can lead to overfitting the data. For this reason we perform several analyses which include different combinations of the (330), (440), (210), and (550) modes, in addition to the dominant (220) mode. Numerical simulations of binary black hole mergers have generally shown these modes to be the strongest~\cite{Borhanian:2019kxt}. Giesler et al.~\cite{Giesler:2019uxc} showed that including overtones of the dominant harmonic allows a quasi-normal mode description of the signal to be used at earlier times, close to merger. We therefore also perform analyses in which we include the first overtone of the dominant harmonic $(\ell mn) = (221)$.  We use a prior range on the (330) amplitude of $[0, 0.5]\,A_{220}$, while for the (221) mode we use $[0, 5]\,A_{220}$; prior ranges for the other modes can be found in the data release accompanying this paper. These choices for the amplitude priors are sufficiently broad that they comfortably include results from numerical simulations~\cite{Borhanian:2019kxt,Giesler:2019uxc}. Other prior choices are possible; e.g., a broader amplitude prior was adopted in Ref.~\cite{Abbott:2020jks}.

We repeat these analyses in $1\,\rm{ms}$ intervals between $t_{\rm ref} + [-9\,\rm{ms},\, 24\,\rm{ms}]$. We use Bayes factors to determine which model is most favored at each time step. For the model that includes the fundamental dominant harmonic (220) and its overtone (221), the Bayes factor is evaluated against the model with only the (220) mode. For models that include the (330) (or other sub-dominant modes), the Bayes factor is evaluated against the stronger of the (220) or (220)+(221) models.

The Bayes factors for the various multimode Kerr models are shown in Fig.~\ref{fig:Bayes_vs_t}. Consistent with the agnostic results, we find strong evidence for the presence of the $(330)$ mode, with the Bayes factor for the $(220)+(330)$ model peaking at \BFthreethree{} at $t_{\rm ref}+6\,$ms. The marginal posterior on the $(330)$ amplitude is peaked away from zero at this time with a value of $A_{330}/A_{220} = \ampThreeThree{}$; see Fig.~\ref{fig:amp_posterior}. The maximum likelihood ringdown waveforms at this time are shown in the Supplemental Material.

A Bayes factor of $\sim 50$ should correspond to a false alarm probability of $0.02$. In Ref.~\cite{Capano:2022zqm} we validate this by repeating the analysis on a large number of simulated signals added to off-source data surrounding GW190521. There, we find that the distribution of Bayes factors in noise matches expectations; e.g., the probability of obtaining a maximized Bayes factor as large as $56$ from noise is $\sim 0.02$. Our quoted Bayes factor is therefore robust against background noise fluctuations.

As can be seen in Fig.~\ref{fig:Bayes_vs_t}, the model that includes the (220), (330), and (210) modes is nearly as strong as the model with just the $(220)$ and $(330)$ mode in it, and is slightly favored at $t_{\rm ref}+18\,$ms, indicating some support for the presence of the (210) mode in addition to the (330). However, the ratio of evidences between the (220)+(330) and the (220)+(330)+(210) models is order unity --- i.e., the data is uninformative as to whether the (210) mode is observable \emph{in addition to} the (330). A model consisting of just the (220) and (210) is disfavored at all times compared to any involving the (330), as can be seen in Fig.~\ref{fig:Bayes_vs_t}. We therefore only claim detection of the (330) mode, and make no claim regarding the observability of the (210) mode.

\begin{figure}
\includegraphics[width=1.0\linewidth]{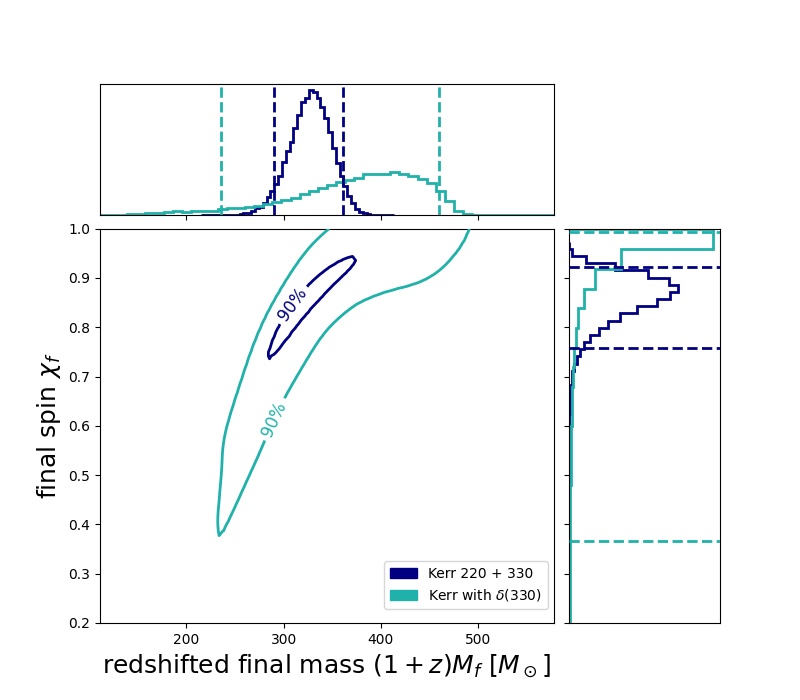}
\caption{Posterior distribution of final redshifted mass $(1+z)M_f$ and dimensionless spin $\chi_f$ measured at $6\,$ms after $t_{\rm ref}$ assuming the identified modes are the (220) and (330) modes of a Kerr black hole. Dashed lines indicate the 90\% credible interval. For the Kerr with $\delta(330)$ results, we use fitting formulae~\cite{Berti:2005ys} to convert the frequency $f_{330}(1+\delta f_{330})$ and damping time $\tau_{330}(1+\delta \tau_{330})$ into mass and spin.}
\label{fig:kerrconsistency}
\end{figure}

Figure \ref{fig:kerrconsistency} shows the redshifted mass and the dimensionless spin of the final black hole, measured with the $(220)+(330)$ Kerr model at $6\,$ms after $t_{\rm ref}$. We find that the remnant black hole has a redshifted mass $(1+z)M_f = \estMf\,\mdot$ and dimensionless spin $\chi_f = \estSf$.

If a quasi-normal model without overtones is used too close to merger, the resulting final mass estimate can be biased toward larger values~\cite{TheLIGOScientific:2016src,Giesler:2019uxc}. We find the final mass estimate to be stable between $6\,$ and $12\,$ms using the $(220)+(330)$ model (see the Supplemental Material for plot). This indicates that by this time the black hole has reached a regime of constant ringdown frequency --- a requirement for the validity of linear-regime, quasi-normal modes.

Given the strong evidence for the presence of the $(330)$ mode at $t_{\rm ref}+6\,$ms, we can perform the classic no-hair theorem test~\cite{Dreyer:2003bv} (see also Ref.~\cite{Gossan:2011ha} for a reformulation in terms of parametric deviations). Here, we keep the dependence of $f_{220}$ and $\tau_{220}$ on $(M_f,\chi_f)$ as in the Kerr solution but introduce fractional deviations $\delta f_{330}$ and $\delta\tau_{330}$ of $f_{330}$ and $\tau_{330}$, respectively. We find good agreement with general relativity. Figure~\ref{fig:kerrconsistency} shows the Kerr black hole mass $M_f$ and dimensionless spin $\chi_f$ associated to the (330) mode frequency $f_{330}(1+\delta f_{330})$ and damping time $\tau_{330}(1+\delta \tau_{330})$ measured at $6\,$ms after $t_{\rm ref}$. Plots of the posterior distributions on the fractional deviations are provided in the Supplemental Material. We constrain the fractional deviation from Kerr to $\delta f_{330} = \dfThreeThreeMain$. The damping time is only weakly constrained, with $\delta\tau_{330} = \dTauThreeThreeMain{}$.

\section{Discussion and conclusions}
\label{sec:conclusions}
%********************************************************

The redshifted final mass of GW190521 measured by the LIGO and Virgo Collaborations (LVC) using a (220) ringdown fit was $(1+z)M_f = 282.2^{+50.0}_{-61.9}\mdot$, or $259.2^{+36.6}_{-29.0}\mdot$ when analyzing the full signal~\cite{Abbott:2020niy}. The low-mass-ratio part of the posterior of Nitz and Capano~\cite{Nitz:2020mga} found $(1+z)M_f \sim 260\mdot$ using the full signal~\cite{Varma:2018mmi,Pratten:2020ceb}. These results are somewhat in tension with the final mass and spin inferred from the ringdown modes found here. 

However, the complete waveform models used in the above analyses may not include all relevant physical effects. This, coupled with the fact that GW190521 has a very short inspiral signal, can lead to systematic errors for parameter estimation. For example, the waveform models used in the LVC analysis and Nitz \& Capano assume quasi-circular orbits, but several studies have indicated that the binary may have been eccentric at merger~\cite{CalderonBustillo:2020odh,Romero-Shaw:2020thy,Gayathri:2020coq}. These studies have also found slightly larger estimates for the binary's total mass, making them more consistent with our estimate for the final mass. Even without eccentricity, the reanalysis in Estelles et al.~\cite{Estelles:2021jnz} using a recalibrated time-domain model found a bimodal distribution for the final mass and spin. One of these modes yields similar estimates for the mass and spin as we obtain here; see Ref.~\cite{Capano:2022zqm} for a more detailed comparison.

The ringdown waveforms used in this paper are simpler and more robust than full inspiral-merger-ringdown models for signals like GW190521, provided they are applied sufficiently late in the post-merger regime.  This argument would tend to favor the estimates derived in this paper for the total mass. Nevertheless, a full resolution of this tension is beyond the scope of this work.

Evidence for overtones of the (220) mode very close to merger were previously found for the events GW150914~\cite{Isi:2019aib} and GW190521\_074359~\cite{Abbott:2020jks} (not to be confused with GW190521), although the strength of the evidence for the overtone is disputed~\cite{Cotesta:2022pci,Isi:2022mhy}. Black hole spectroscopy tests showed consistency with the Kerr hypothesis for these events~\cite{Isi:2019aib, Abbott:2020jks}. However, the resulting constraints were weaker than what we find with the (330) mode here. Furthermore, while there is strong numerical evidence for the presence of ringdown overtones close to the merger~\cite{Giesler:2019uxc}, a number of theoretical questions remain as to the validity of a quasi-normal description of the black hole close to merger~\cite{Nollert:1996rf,Nollert:1998ys,Daghigh:2020jyk,Qian:2020cnz,Jaramillo:2020tuu}.

The true nature of the gravitational wave event GW190521 has been the subject of much speculation~\cite{Sakstein:2020axg,Safarzadeh:2020vbv,CalderonBustillo:2020srq}. The interpretation of GW190521 as a head-on collision of two highly spinning Proca stars~\cite{CalderonBustillo:2020srq} predicts the presence of a $(200)$ mode~\cite{Palenzuela:2006wp}. We do not find evidence for such a mode. Additionally, the high-mass, multiple-mode ringdown signal observed here does not agree with the scenario of a very massive star collapsing to a black hole of mass $\sim 50 M_\odot$ and an unstable massive disk~\cite{Shibata:2021sau}.

Expectations based on population models were that black hole ringdown signals with multiple modes were unlikely to be observed with the Advanced LIGO and Virgo detectors~\cite{Berti:2016lat,Cabero:2019zyt} (although those population models did not include massive binaries). However, Forteza et al.~\cite{JimenezForteza:2020cve} predicted that for even moderately asymmetric binaries (with mass ratios $\gtrsim 1.2)$, the $(330)$ mode would be the best observable mode. They further predicted that the (330) mode's frequency could be constrained to the $\sim10\%$ level if the ringdown SNR is $\gtrsim8$. Our results are remarkably consistent with this prediction: we get a ringdown SNR for GW190521 of $\sim 12$, and we constrain the $(330)$ frequency to be within $\sim10\%$ of the expected value from General Relativity.

In summary, we have shown that GW190521 displays a distinct subdominant mode and that this mode is consistent with the $(330)$ ringdown mode of a Kerr black hole.

\section*{Data \& code availability}

Posterior data samples and data necessary to reproduce the figures are available at \url{https://github.com/gwastro/BH-Spectroscopy-GW190521}. The gravitational-wave data used in this work were obtained from the Gravitational Wave Open Science Center (GWOSC) at \url{https://www.gw-openscience.org}.

All software used in this analysis is open source. Bayesian inference was performed with the PyCBC library, available at \url{https://github.com/gwastro/pycbc}. Configuration files used to perform all analyses can be found at \url{https://github.com/gwastro/BH-Spectroscopy-GW190521}. Spheroidal harmonics, Kerr frequencies, and Kerr damping times generated using pykerr, available at \url{https://github.com/cdcapano/pykerr}.

\begin{acknowledgments} 
\section*{Acknowledgments}

The authors thank Ofek Birnholtz, Jose Luis Jaramillo, Reinhard Prix, Bruce Allen, Evan Goetz, Saul Teukolsky, Maximiliano Isi, Juan Calder{\'on}-Bustillo, Abhay Ashtekar and Bangalore Sathyaprakash for useful discussions and Xisco Jim{\'e}nez Forteza for a careful reading of this manuscript. We thank also the Atlas Computational Cluster team at the Albert Einstein Institute in Hanover for assistance. MC acknowledges funding from the Natural Sciences and Engineering Research Council of Canada (NSERC).
This research has made use of data obtained from the Gravitational Wave Open Science Center (https://www.gw-openscience.org/ ), a service of LIGO Laboratory, the LIGO Scientific Collaboration and the Virgo Collaboration. LIGO Laboratory and Advanced LIGO are funded by the United States National Science Foundation (NSF) who also gratefully acknowledge the Science and Technology Facilities Council (STFC) of the United Kingdom, the Max-Planck-Society (MPS), and the State of Niedersachsen/Germany for support of the construction of Advanced LIGO and construction and operation of the GEO600 detector. Additional support for Advanced LIGO was provided by the Australian Research Council. Virgo is funded, through the European Gravitational Observatory (EGO), by the French Centre National de Recherche Scientifique (CNRS), the Italian Istituto Nazionale di Fisica Nucleare (INFN) and the Dutch Nikhef, with contributions by institutions from Belgium, Germany, Greece, Hungary, Ireland, Japan, Monaco, Poland, Portugal, Spain.
\end{acknowledgments}

%\printbibliography

%\noindent \textbf{Supplementary materials:} \\
%Materials and Methods\\
%Figures~\ref{supfig:data} -- \ref{supfig:delta221}

\clearpage
\section*{Supplemental Material}

\renewcommand\thefigure{S.\arabic{figure}}
\setcounter{figure}{0}

%\section*{Materials and Methods}
%********************************************

\subsection*{The ringdown signal model}

In the quasi-normal mode (QNM) spectrum of a perturbed Kerr black hole, the allowed frequencies $f_{\lmn} = \omega_{\lmn}/(2\pi)$ and damping times $\tau_{\lmn}$ are labeled by three integers $\ell = 2,3,\dots$, $-\ell \leq m \leq \ell$, and $n=0,1,2,\ldots$. These can be combined together in a complex frequency $\Omega_{\lmn}=\omega_{\lmn }+i/\tau_{\lmn}$ such that the ringdown signal of a perturbed Kerr black hole can be expressed as a sum of damped sinusoids:
\begin{equation}
\label{linear_spectrum}
h_+ + i h_\times = {\displaystyle \frac{M_f}{D_L}} \sum_{\lmn} {}_{_{-2}}S_{\l m n} (\iota, \varphi; \chi_f) 
A_{\lmn} e^{i(\Omega_{\lmn}t +  \phi_{\lmn})} \, ,
\end{equation}
where $h_+$ and $h_\times$ are the plus and cross polarizations of the gravitational wave, $M_f$ is the mass of the black hole in the detector frame and $D_L$ is the luminosity distance to the source. The functions ${}_{_{-2}}S_{\l m n} (\iota, \varphi; \chi_f)$ are the spin-weighted spheroidal harmonics of spin weight $-2$, which depend on the inclination angle $\iota$ between the black hole spin and the line-of-sight from the source to the observer, the azimuth angle $\varphi$ between the black hole and the observer, and the dimensionless spin $\chi_f$ of the black hole.

The complex QNM frequencies $\Omega_{\lmn}$ can be determined from the Teukolsky equation~\cite{Teukolsky:1972my,Leaver:1985ax}. According to the no-hair theorem, the frequencies and damping times are determined by the mass $M_f$ and spin $\chi_f$ of the black hole, with $\chi_f \in (-1, 1)$. Positive (negative) spin means the perturbation is co(counter)-rotating with respect to the black hole. The amplitudes $A_{\lmn}$ and phases $\phi_{\lmn}$ depend on the initial perturbation and take different values for different $\lmn$ modes. 

For a given $\ell$ and $n$, the $+m$ and $-m$ modes are related to each other by $\omega_{\ell -m n} = -\omega_{\ell m n}$ and $\tau_{\ell -m n} = \tau_{\ell m n}$~\cite{Berti:2005ys}. Furthermore, if the initial perturbation is symmetric under reflection at the equatorial plane, the amplitude and phase of the $\pm m$ modes are related to each other by $A_{\ell-mn}e^{i\phi_{\ell-mn}} = (-1)^{\ell} A_{\ell m n}e^{-i\phi_{\ell m n}}$. Numerical simulations of black hole binaries have shown that it is difficult to break reflection symmetry in the ringdown~\cite{Berti:2007fi}. To simultaneously sum over the $\pm m$ modes for a given $\ell n$, we parameterize the waveform as
\begin{align*}
 h_{\ell |m| n} &= A^0_{\ell|m|n}e^{-t/\tau_{\ell|m|n}} \times \\
&[{}_{-2}S_{\ell mn}(\iota,\varphi) A^{(+)}_{\ell|m|n} e^{i(\omega_{\ell|m|n} t +\phi_{\ell|m|n})} \\
&\,+ {}_{-2}S_{\ell-mn}(\iota,\varphi)A^{(-)}_{\ell|m|n} e^{-i(\omega_{\ell|m|n} t + \phi_{\ell|m|n})}],
\end{align*}
where $A^0_{\ell |m| n}$ is the intrinsic amplitude of the ($\ell m n$) mode and
\begin{align}
    A^{(+)}_{l|m|n} &= \sqrt{2}\cos(\pi/4 + \Delta \beta_{l|m|n}) \\
    A^{(-)}_{l|m|n} &= \sqrt{2}\sin(\pi/4 + \Delta \beta_{l|m|n}) e^{i (l\pi + \Delta \phi_{l|m|n})}.
\end{align}
If the parameters $\Delta \beta_{\ell m n}$ and $\Delta \phi_{\ell m n}$ are both zero, the waveform reduces to the case found for reflection symmetry.

When using the Kerr model, we do two analyses at several times: one with reflection symmetry, for which $\Delta \beta_{\ell m n}$ and $\Delta \phi_{\ell m n}$ are both set to zero, and one in which a common $\Delta \beta$ and $\Delta \phi$ for all modes are allowed to vary uniformly between $[-\pi/4, \pi/4)$ and $[-\pi,\pi)$, respectively.

We find the analysis with reflection symmetry is slightly favored by the data. For example, the maximum (330) Bayes (which also peaks at $t_\mathrm{ref}+6\,$ms) is $42$ when reflection symmetry is broken, compared to 56 when reflection symmetry is assumed (as discussed in the main text). We also find that we recover consistent values for parameters --- final mass, spin, and (330) amplitude --- for both models, and that $\Delta \beta$ and $\Delta \phi$ are centered on zero in the model without reflection symmetry. We therefore only report results from the model with reflection symmetry in the main text.

In all analyses we fix the azimuthal angle $\varphi = 0$, as it is degenerate with the modes' initial phases. To obtain the frequency and damping times for a given mass and spin we use tabulated values from Berti et al.~\cite{Berti:2005ys}, which we interpolate using a cubic spline. For the spheroidal harmonics we use tabulated values of the angular separation constants (also from Berti et al.~\cite{Berti:2005ys}) and solve the recursion formula given in Leaver~\cite{Leaver:1985ax}. Our code for doing this is publicly available on GitHub~\cite{pykerr}.

For the agnostic analysis, we do not assume any mode corresponds to any particular $\ell m n$. We therefore use arbitrary complex numbers $X_{\ell \pm m n} = e^{i \psi_{\ell \pm m n}}$ in place of the ${}_{-2}S_{\ell \pm m n}$. Here, the $\psi_{\ell \pm m n}$ are allowed to vary uniformly in $[0, 2\pi)$. We vary a common $\Delta \beta$ parameter, but fix the $\Delta \phi$ parameter to zero, as it is degenerate with the $X_{\ell m n}$.

\subsection*{Computational analysis methods}

\renewcommand{\vec}[1]{{\mathbf{#1}}}
\newcommand{\net}{\ensuremath{\mathrm{net}}}
\newcommand{\data}{\ensuremath{s}}
\newcommand{\vdata}{\ensuremath{\vec{\data}}}
\newcommand{\noise}{\ensuremath{n}}
\newcommand{\vnoise}{\ensuremath{\vec{\noise}}}
\newcommand{\signal}{\ensuremath{h}}
\newcommand{\vsignal}{\ensuremath{\vec{\signal}}}
\newcommand{\covmat}{\ensuremath{C}}
\newcommand{\vcovmat}{\ensuremath{\vec{C}}}
\newcommand{\transpose}{\ensuremath{\mathsf{T}}}
\newcommand{\numdet}{\ensuremath{K}}
\newcommand{\ip}[2]{\left< #1, #2 \right>}
\newcommand{\psd}{\ensuremath{S_n}}

Standard parameter estimation with gravitational waves begins with Bayes' theorem. Given some data \vdata{} and a signal model $h$ that depends on some set of parameters $\vec{\lambda}$, we wish to know the posterior probability density function $p(\vec{\lambda}|\vdata, h)$. Applying Bayes' theorem we have
\begin{equation*}
    p(\vec{\lambda}|\vdata, h) = \frac{1}{Z} p(\vdata|\vec{\lambda}, h) p(\lambda|h),
\end{equation*}
where $p(\vdata|\vec{\lambda}, h)$ is the likelihood function, $p(\lambda|h)$ is the prior, and $Z$ is a normalization constant known as the evidence. Estimates on a single parameter are obtained by marginalizing the posterior over all other parameters; marginalizing over all parameters yields the evidence. Taking the ratio of evidences $Z_A/Z_B$ for two different signal models yields the ``Bayes factor''. If our prior belief for the validity of the two models is the same, the Bayes factor gives the odds that model A is favored over model B. Using a scale by Kass and Raftery~\cite{Kass:1995loi}, a Bayes factor greater than 3.2 is considered ``substantial'' evidence in favor of model A; greater than 10 is ``strong'' evidence; greater than 100 is ``decisive''.

Evaluating the posterior requires a likelihood function $p(\vdata|\vec{\lambda}, h)$. Consider a gravitational-wave detector, which we sample every $\Delta t$ seconds over a time $T$ to obtain $N = \lceil T/\Delta t \rceil$ time-ordered samples $\vdata = \{\data_0, ..., \data_{N-1}\}$. A network of $\numdet{}$ detectors sampled in this way will produce a set of samples $\vdata_{\net} = \{\vdata_1, ..., \vdata_{\numdet}\}$. To obtain a likelihood function we first consider the hypothesis that the set of samples only contain noise $p(\vdata_{\net}|\noise) = p(\vnoise_{\net})$.

In gravitational-wave astronomy it is common to assume that, in the absence of a signal, the detectors output stochastic Gaussian noise that has zero mean and is independent across detectors. Under this assumption the probability density function describing the network of time-ordered noise samples $\vnoise_{\net}$ is a product of $\numdet{}$ $N-$dimensional multivariate normal distributions,
\begin{equation}
\label{eqn:pnoise}
p(\vnoise_{\net}) = \frac{\exp\left[
    -\frac{1}{2}\sum_{d=1}^{\numdet{}} \vnoise_{d}^{\top}\vcovmat_{d}^{-1} \vnoise_{d}\right]}    {\sqrt{(2\pi)^{N\numdet{}} \prod_{d=1}^{\numdet{}}\det \vcovmat_{d}}}.
\end{equation}
Here, $\vcovmat_{d}$ is the covariance matrix of the noise in detector $d$. In order to evaluate this function it is necessary to know what the inverse of the covariance matrix is.

If we assume that a detector's noise is wide-sense stationary and ergodic, then its covariance $\vcovmat$ is a symmetric Toeplitz matrix with elements given by the autocorrelation function of the data. If the autocorrelation function goes to zero in some finite amount of time that is less than $T/2$ (for the LIGO and Virgo detectors, this typically happens within a few seconds), then the covariance matrix is asymptotically equivalent to a circulant matrix. The inverse of the covariance matrix can then be well-approximated by that of the equivalent circulant matrix. 
All circulant matrices have the same eigenvectors, $e^{-2\pi i k p/N}/{\sqrt{N}}$~\cite{gray:2006}.
Solving for the eigenvalues yields an analytic expression for $\vcovmat^{-1}$: the $j,k$-th element is the discrete inverse Fourier transform of $1/\psd$ evaluated at the $k-j$ time step,
\begin{equation}
    \covmat^{-1}[j,k] \approx 2 \Delta t^2 \mathcal{F}^{-1}(\psd^{-1})[k-j],
\label{eqn:invcov}
\end{equation}
where $\psd$ is the power spectral density of the detector's noise. Substituting this back into Eq.~\eqref{eqn:pnoise}, yields a canonical likelihood function for the noise~\cite{Finn:1992wt},
\begin{equation}
    p(\vdata_{\net} | \noise) \propto \exp\left[-\frac{1}{2}\sum_{d=1}^{K} \ip{\vnoise_d}{\vnoise_d}\right].
\label{eqn:likelihood}
\end{equation}
Here, the inner product $\ip{\cdot}{\cdot}$ is defined as
\begin{equation}
\ip{\vec{u}_d}{\vec{v}_d} \equiv 4 \Re \left\{\frac{1}{T} \sum_{p=1}^{\lfloor \left( N-1 \right)/2 \rfloor} \frac{\tilde{u}_d^{*}[p]\tilde{v}_d[p]}{\psd^{(d)}[p]} \right\},
\label{eqn:innerprod}
\end{equation}
where $\tilde{u}$ is the discrete Fourier transform of the time series $\vec{u}$ and $^*$ means complex conjugation\footnote{This definition excludes the DC-component $p=0$ for odd $N$, and both DC-component and Nyquist-frequency component $p=N/2$ for even $N$, which can be treated separately or may be negligible~\cite{Allen:2005fk}.}.

The signal hypothesis is that the data consists of the signal plus the noise. The likelihood function $p(\vdata|\vec{\lambda}, h)$ is therefore Eq.~\eqref{eqn:likelihood} with the $\vnoise_d$ replaced by the residuals $\vdata_d - \vsignal_d$. However, this assumes that $\vsignal$ is an accurate model of the signal across the entire observation time $T$. As stated above, quasi-normal modes only model the gravitational wave from a binary black hole after the merger, when the two component black holes have formed a single, perturbed black hole. Performing Bayesian inference using quasi-normal modes as the signal model therefore requires excising times from the data when the ringdown prescription is not valid. In other words, instead of considering the full set of time samples $\vdata = \{\data_0, ..., \data_{N-1}\}$, we wish to only evaluate the truncated set $\vdata_{tr} = \{\data_0, ..., \data_a, \data_{a+M}, ..., \data_{N-1}\}$, with $M > 1$. The data between the time steps $[a,a+M)$ is said to be ``gated''. 

The probability density function of the truncated noise is still a multivariate normal distribution (excising dimensions from a multivariate normal is equivalent to marginalizing over those dimensions), and so Eq.~\eqref{eqn:pnoise} still applies. The challenge is that the covariance matrix of the truncated noise $\vcovmat_{tr}$ is no longer Toeplitz. Its eigenvectors can no longer be approximated by that of a circulant matrix, and so the expression for the likelihood Eq.~\eqref{eqn:likelihood} is no longer valid. The inverse of the covariance matrix needs to be found by other means. One possibility is to numerically invert the covariance matrix. However, this is numerically unstable due to the large dynamic range of the matrix's elements, and computationally impractical for observation times of more than a $\sim$second. Instead, we use ``gating and in-painting'' to find the likelihood of the truncated time series. This was applied to the problem of matched filtering in Zackay et al.~\cite{Zackay:2019kkv}. Here we apply it to parameter estimation.

Define $\vnoise' = \vnoise_g + \vec{x}$, where $\vnoise_g$ is the noise with the gated times $t\in [a, a+M)\Delta t$ zeroed out, and $\vec{x}$ is a vector that is zero everywhere except in the gated times. If the non-zero elements of $\vec{x}$ are such that $(\vcovmat^{-1} \vnoise')[k] = 0$ for all $k \in [a, a+M)$, then $\vnoise'^\transpose \vcovmat^{-1} \vnoise'$ will  be the same as the truncated version $\vnoise_{tr}^\transpose \vcovmat_{tr}^{-1} \vnoise_{tr}$. Our aim is to solve the equation $\vcovmat^{-1}(\vnoise_g + \vec{x}) = \vec{0}$ in the gated region. Since $\vec{x}$ is zero outside of the gated region, $\vcovmat^{-1}\vec{x}$ only involves  the $[a, a+M)$ rows and columns of $\vcovmat^{-1}$, which form an $M \times M$ Toeplitz matrix [cf. Eq.~\eqref{eqn:invcov}]. We therefore solve for $\vec{x}$ such that
\begin{equation}
    \overline{\vcovmat^{-1}}\,\overline{\vec{x}} = -\overline{\vcovmat^{-1}\vnoise_g},
\label{eqn:gatecondition}
\end{equation}
where the overbar indicates the $[a, a+M)$ rows (and columns) of the given vector (matrix). This can be solved numerically using a Toeplitz solver~\cite{Virtanen:2019joe,Harris:2020xlr}. Adding $\vec{x}$ to the gated noise (``in painting'') will then yield the same result as if we had truncated the noise and the covariance matrix. 

Note that if the gate spans the entire beginning of the data segment, the truncated covariance matrix $\vcovmat_{tr}$ is Toeplitz, and so could be inverted numerically using a Toeplitz solver. This is the method used by Isi et al.~\cite{Isi:2019aib}. The advantage of using in-painting is that it involves solving for an $M\times M$ matrix rather than an $(N-M)\times (N-M)$ matrix. Gating and in-painting also have other applications beyond what we use it for here, such as excising glitches from data when doing parameter estimation.

To evaluate the likelihood for a signal, we use $\vnoise_g = \vdata_g - \vsignal_g$ (i.e., the residual with the gated region zeroed out) in Eq.~\eqref{eqn:gatecondition} and solve for $\vec{x}$. We can then use $\vec{x} + \vdata_g - \vsignal_g$ in the standard likelihood, Eq.~\eqref{eqn:likelihood}. We do not attempt to normalize the likelihood, which would involve finding the determinant of the truncated covariance matrix. For this reason, we calculate and report Bayes factors by comparing models that start at the same time offset from $t_{\rm ref}$, for which the determinant cancels.

We use the open source PyCBC Inference library for performing Bayesian inference~\cite{pycbcgithub,Biwer:2018osg}, to which we have added the gated likelihood described above. For all analyses we use a gate of two seconds, ending at the start time of the ringdown. For sampling the parameter space we use the dynesty nested sampler~\cite{speagle:2019} and we numerically marginalize over polarization to speed up convergence. To better estimate the Bayes factor of the (220)+(330) model, we repeat the (220), (220)+(221), and (220)+(330) analyses 11 times, with one of the 11 using twice as many live points as the other 10. Bayes factors are averaged over the 11 analyses to reduce the uncertainty; parameter estimates are quoted using the run with the larger number of live points.

We use data for the event GW190521 made publicly available by the Gravitational Wave Open Science Center~\cite{Abbott:2019ebz}. 
We fix the sky location to the values given by the maximum likelihood result of Nitz \& Capano~\cite{Nitz:2020mga}, although we have obtained similar results using the LVC's maximum likelihood sky location~\cite{Abbott:2020mjq}. When we use that sky location the (220)+(330) Bayes factor peaks at the same time as reported in our work, with a value of $\sim45$.

We use a geocentric GPS reference time of $t_{\rm ref} =1242442967.445$~\cite{Nitz:2020mga}. 
With the sky location used in our analyses, this corresponds to the detector GPS reference times $1242442967 + 0.4259$ at LIGO Hanford, $+0.4243$ at LIGO Livingston and $+0.4361$ at Virgo. Credible intervals in the text are quoted to $90$\%.

\subsection*{Mass ratio estimate}

\begin{figure}
\centering
\includegraphics[width=0.98\columnwidth]{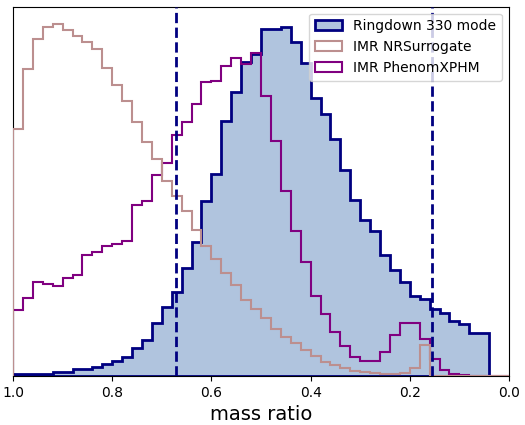}
\caption{Posterior distribution of the estimated mass ratio of the binary, $m_2/m_1 < 1$. This is obtained by applying the numerical fits from Ref.~\cite{Borhanian:2019kxt} that relate the (330) amplitude to mass ratio~\cite{Borhanian:2019kxt} to our measured amplitude ratio (Fig.~\ref{fig:amp_posterior}). Vertical dashed lines indicate the 90\% credible interval. We assume a prior uniform in mass ratio between $1/25$ -- $1$ when applying the fits to the amplitude posterior. For comparison, we also show the mass ratio obtained from the full signal from Ref.~\cite{Nitz:2020mga} using NR Surrogate and from Ref.~\cite{Nitz:2021uxj} using PhenomXPHM models assuming the same prior.}
\label{fig:mass-ratio}
\end{figure}

The detection of a (330) mode indicates that the progenitor black holes in GW190521 had asymmetric masses, since equal-mass binaries are not expected to excite the (330) mode. Numerical fits~\cite{Borhanian:2019kxt} provide the relation between the amplitude of the (330) mode and the ratio of the initial black hole masses $m_1$ and $m_2$ for quasi-circular, aligned spin binaries. We find $m_2/m_1=\qthreethree$ from the (330) amplitude measured by the $(220)+(330)$ Kerr model at $6\,$ms after $t_{\rm ref}$. Figure \ref{fig:mass-ratio} shows the posterior distribution on the amplitude of the (330) mode and on the corresponding mass ratio. We caution that this result for the mass ratio depends on the model of Ref.~\cite{Borhanian:2019kxt}, which only considered quasi-circular, aligned-spin binaries. The relationship between the (330) amplitude and the binary's mass ratio may differ if GW190521 was precessing or had appreciable eccentricity at merger.

\subsection*{No-hair test using the 221 overtone}

\begin{figure*}
\centering
\includegraphics[width=\textwidth]{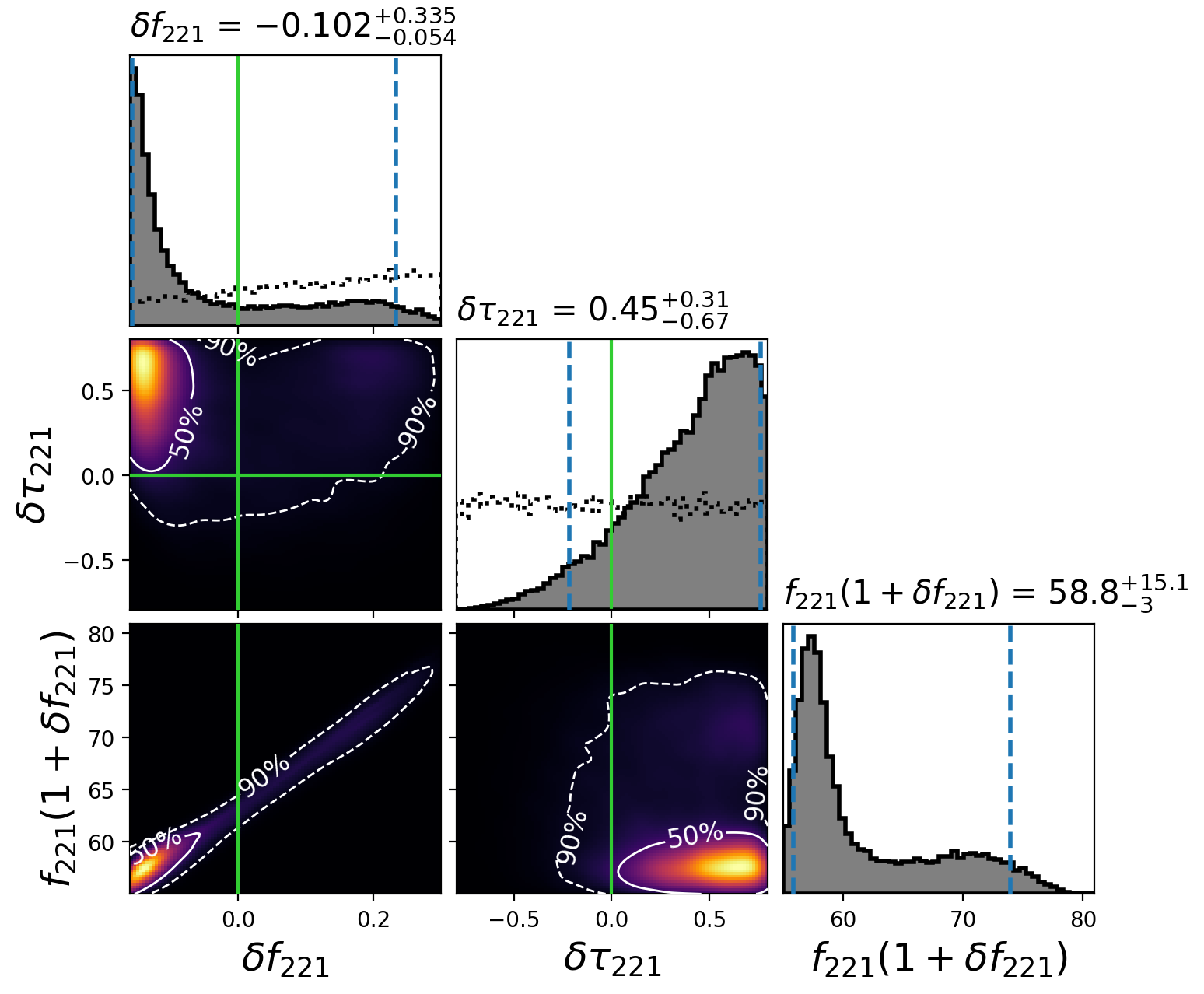}
\caption{Posterior on the deviation from Kerr of the (221) frequency $\delta f_{221}$ and damping time $\delta \tau_{221}$, as well as the resulting modified $(221)$ frequency, using a model in which we include the (220)+(221) modes at $t_{\rm ref}-7\,$ms. Quoted values are the median and 90\% credible interval, which is indicated by the dashed vertical lines. The fractional deviations are expected to be zero for a Kerr black hole (indicated by the green lines). We use a prior (black dotted lines) that is uniform over $\delta f_{221}\in[-0.16, 0.3)$, with the constraint that $f_{221}(1+\delta f_{221}) > 55$\,Hz. This constraint is used to try to exclude additional noise that is present in the Hanford and Virgo detectors at $\sim50\,$Hz. For the damping time we use a prior that is uniform over $\delta \tau_{221} \in [-0.8, 0.8)$. Despite the tighter prior constraints than that used for the $(330)$ mode, we obtain a larger $90\%$ credible interval on both $\delta f_{221}$ and $\delta \tau_{221}$. The posterior also peaks toward the prior boundaries. This may be due to the noise at low frequency, or may indicate that the signal is not fully captured by a sum of quasi-normal modes at this time.}
\label{supfig:delta221}
\end{figure*}

Given the evidence we find for the $(221)$ model at $t_{\rm ref} - 7\,$ms, we also perform a no-hair theorem test on the $(221)$ overtone at this time. The results are shown in Fig.~\ref{supfig:delta221}. We find poor constraints, with the posterior peaking away from the expected Kerr result. We find similar results when performing the same analysis on simulated signals that do not violate general relativity; see Ref.~\cite{Capano:2022zqm} for details. This may indicate that even if overtones are present, they may not yield a complete description of the signal at merger. More study is needed to confirm this.

\newpage
\onecolumngrid
\section*{Additional figures}

\begin{figure*}[h]
\centering
\includegraphics[width=0.45\columnwidth]{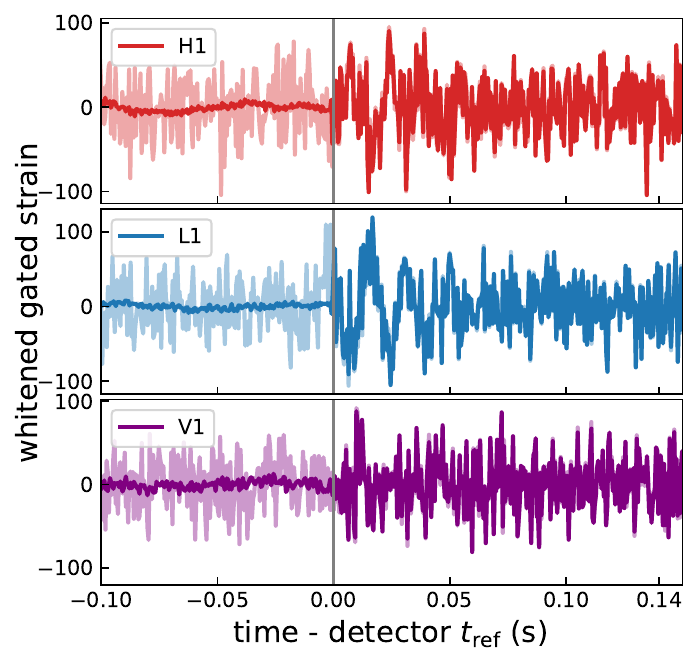} \hfill
\includegraphics[width=0.45\columnwidth]{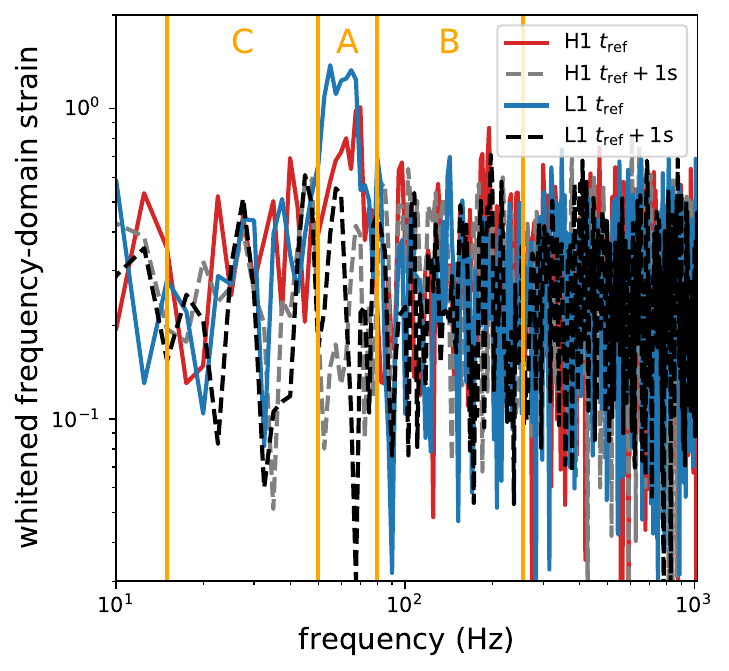}
\caption{\textit{Left:} Whitened data in each detector. Dark traces show the data gated at the reference time $t_{\rm ref}$ in each detector, which is indicated by the gray vertical lines. \textit{Right:} Frequency domain representation of the Hanford and Livingston data shown in the left panel. Also shown is the frequency domain representation at an off-source time, one second later. The signal is clearly visible in the LIGO time-domain data, and is seen as a spike in the frequency domain data between $\sim60-70\,$Hz. The primary frequency bin (``A'') boundaries were set to isolate this spike. Frequencies below (region ``C'') and above (region ``B'') were searched for additional QNMs in the agnostic analysis.}
\label{supfig:data}
\end{figure*}

\begin{figure*}[!hb]
\centering
\includegraphics[width=0.45\columnwidth]{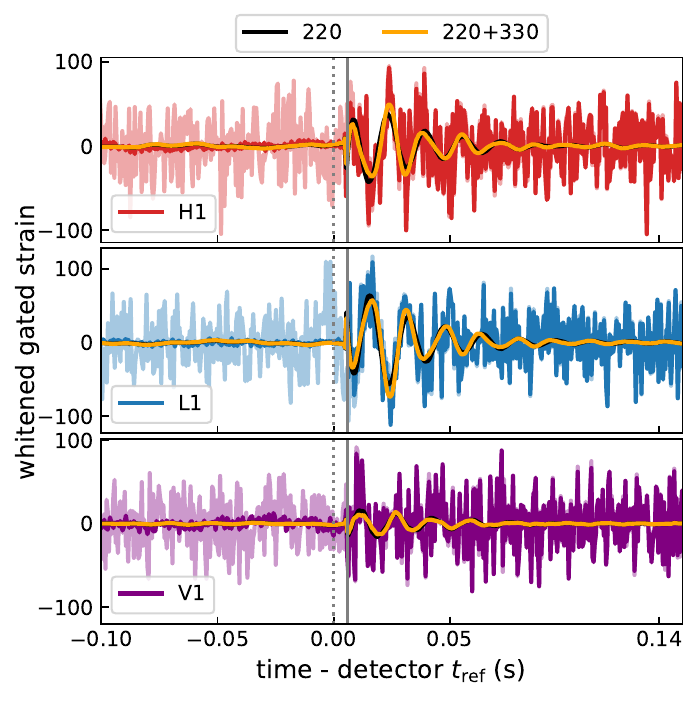}
\caption{Whitened data in each detector, with a gate applied at $6\,$ms (gray lines) after the detector reference time $t_{\rm ref}$ (gray dotted lines). Semi-transparent traces show the whitened data without the gate. Plotted are the maximum likelihood waveforms using just the (220) mode (black) and the (220) plus (330) mode (orange).}
\label{supfig:maxL_data}
\end{figure*}

\begin{figure*}
\centering
\includegraphics[height=0.21\textheight]{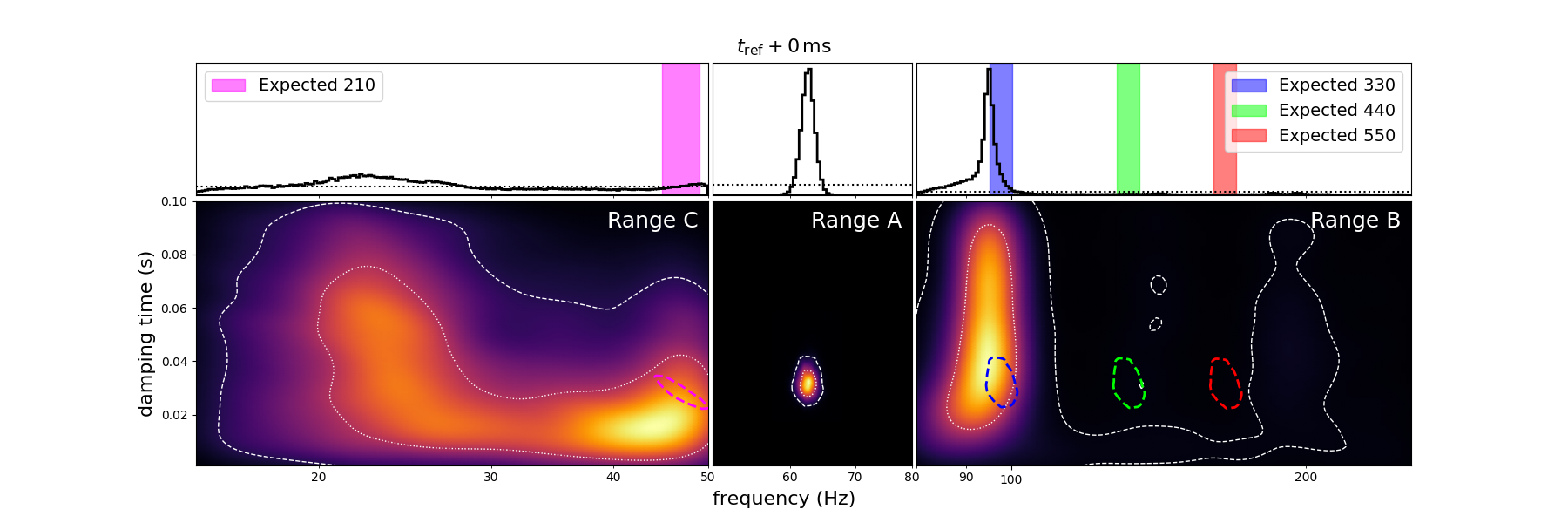} \\
\includegraphics[height=0.21\textheight]{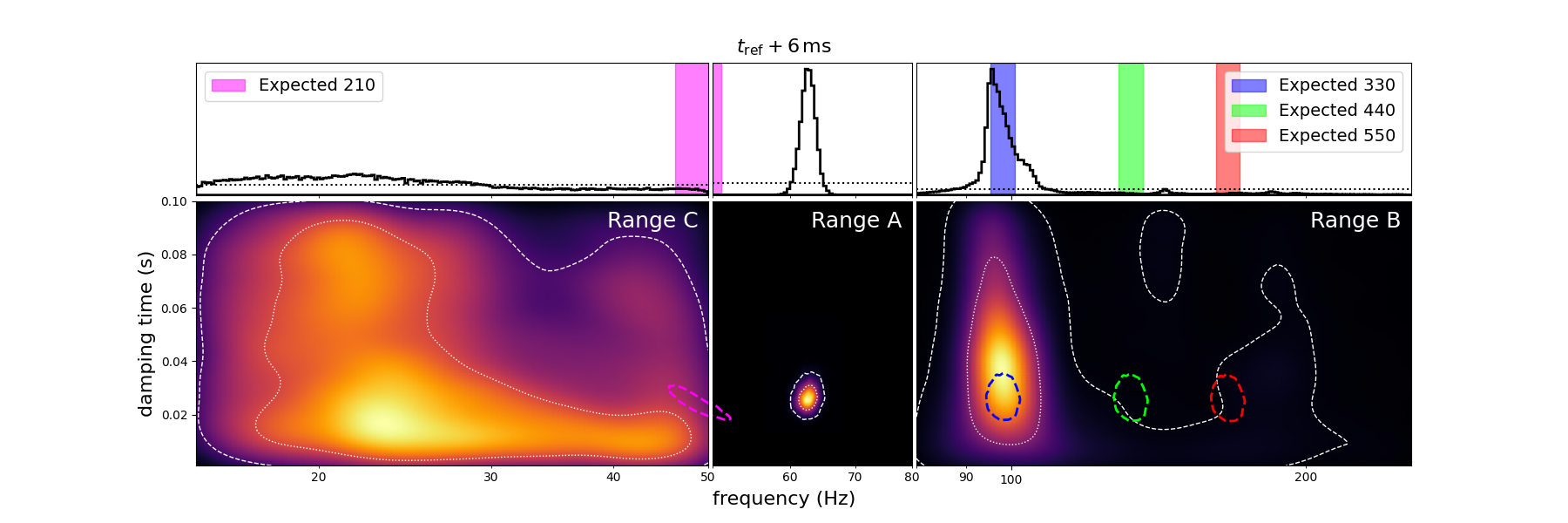} \\
\includegraphics[height=0.21\textheight]{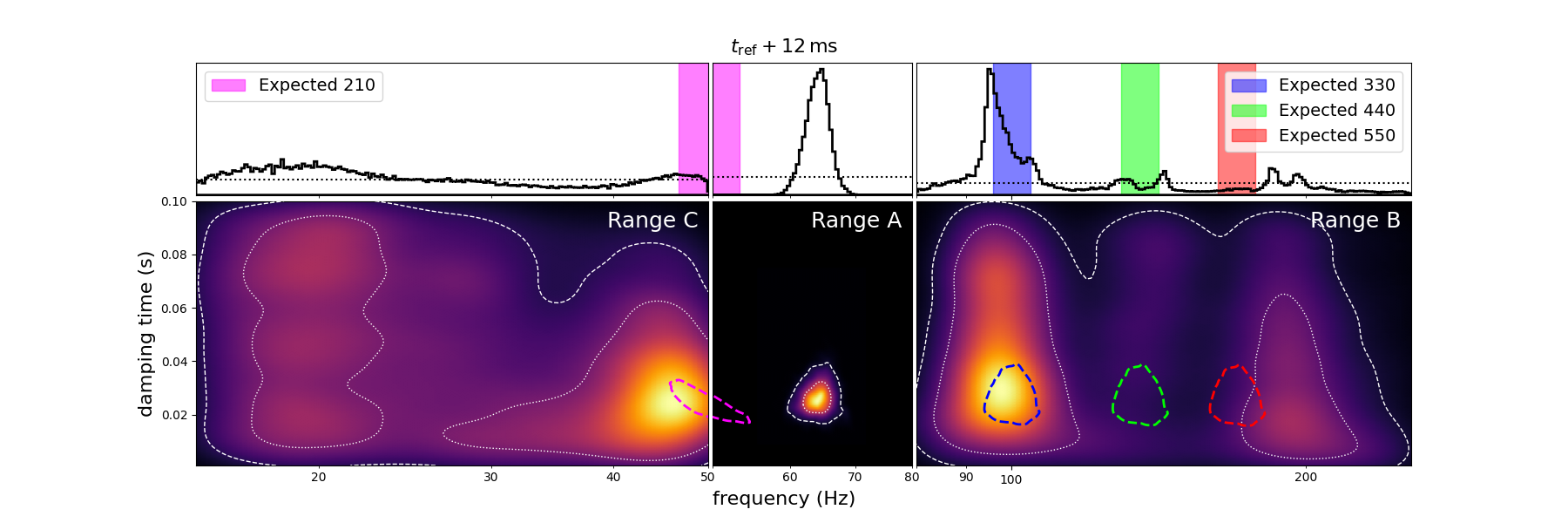} \\
\includegraphics[height=0.21\textheight]{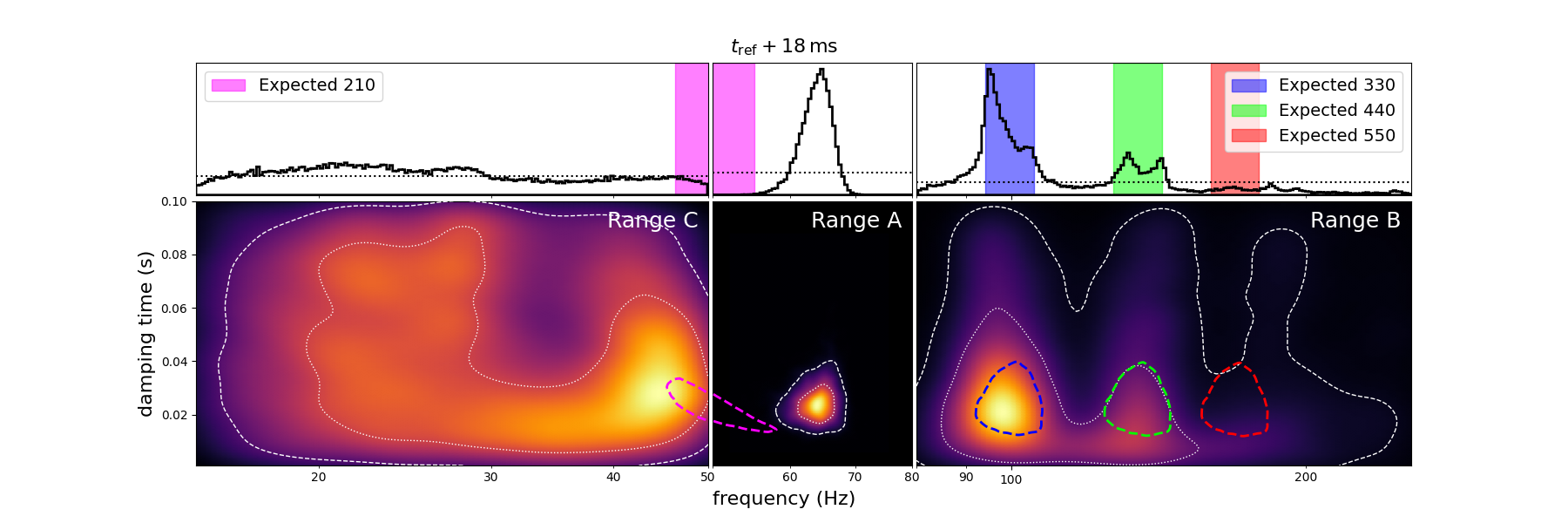}
\caption{Spectra plots at 0, 6, 12, and $18\,$ms showing marginal posterior distributions from frequency ranges A, B, and C in the agnostic analysis. Also shown are the expected regions for the (210) (magenta), (330) (blue), (440) (green), and (550) (red) modes, assuming the peak in region A is the (220) mode. A mode is clearly visible in range A and in range B; C is consistent with noise (note the small difference between the posterior and prior in range C). Both the mode in range A and the mode in range B shift to higher frequency between 0 and $6\,$ms, then remain largely constant while the posterior widens. A secondary peak consistent with the (440) mode is noticeable at $18\,$ms, but the signal is weak at this point.}
\label{supfig:agnostic}
\end{figure*}

\begin{figure*}
\centering
\includegraphics[width=0.75\textwidth]{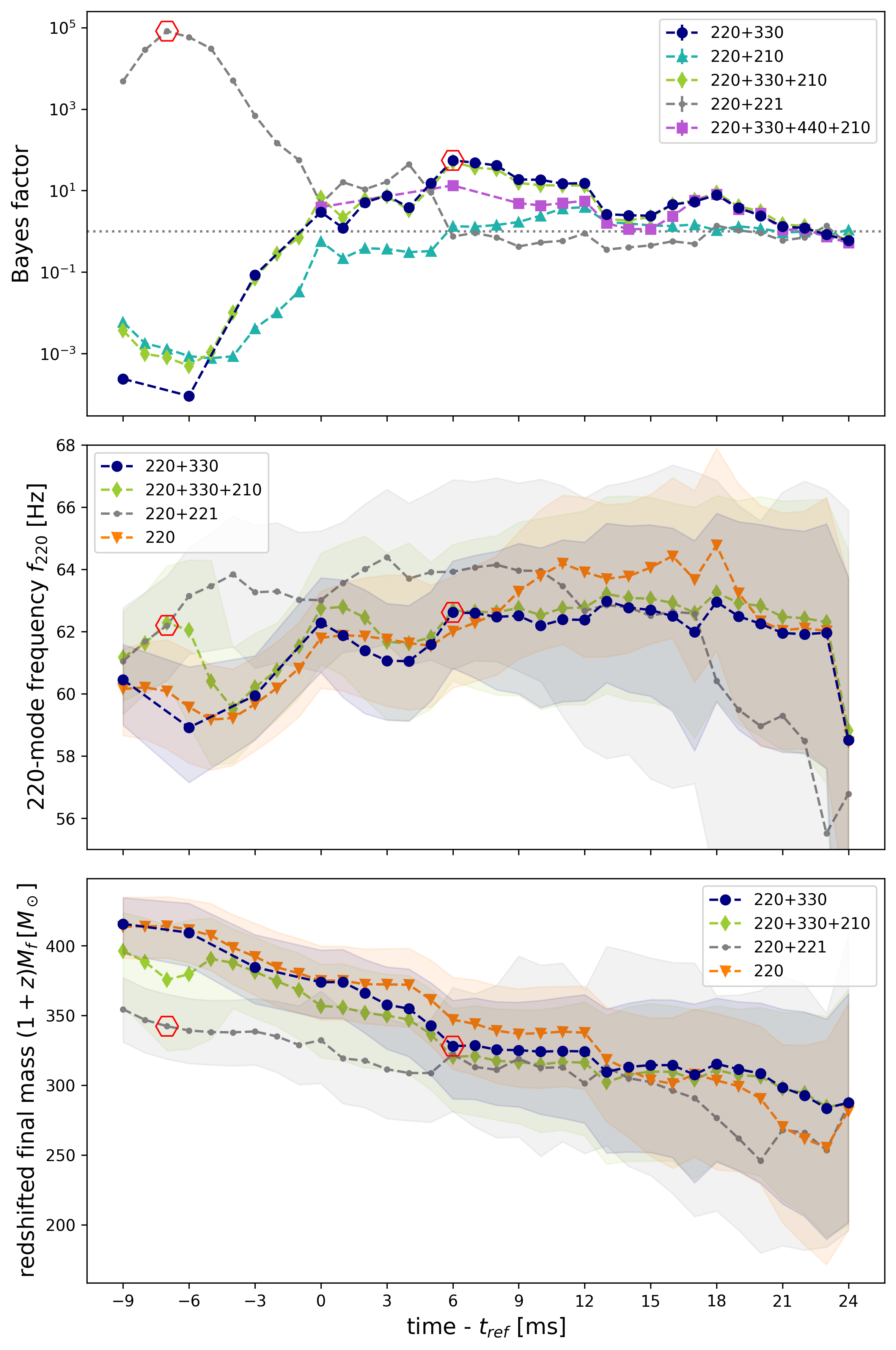}
\caption{Bayes factors as shown in Fig.~\ref{fig:Bayes_vs_t} for an extended range of times. (\textit{Top}) Bayes factor of models with the indicated modes compared to the stronger of the (220) or the (220) + (221) modes model. The Bayes factor for the (220) + (221) model is calculated against the (220)-only model. We additionally show the results for a model that includes the (440) mode (purple line) here. (\textit{Center}) Median values for the frequency of the (220) mode for the model with the indicated modes. The shading shows the 90\% credible interval.
(\textit{Bottom}) Median values and 90\% credible intervals for the redshifted final mass.
Hexagons mark where no-hair tests are performed for the (220) + (330) model and for the (220) + (221) model.
All values are shown for start times of the analysis relative to the reference time $t_{\rm ref}$.}
\label{supfig:BF}
\end{figure*}

\begin{figure*}
\centering
\includegraphics[width=\textwidth]{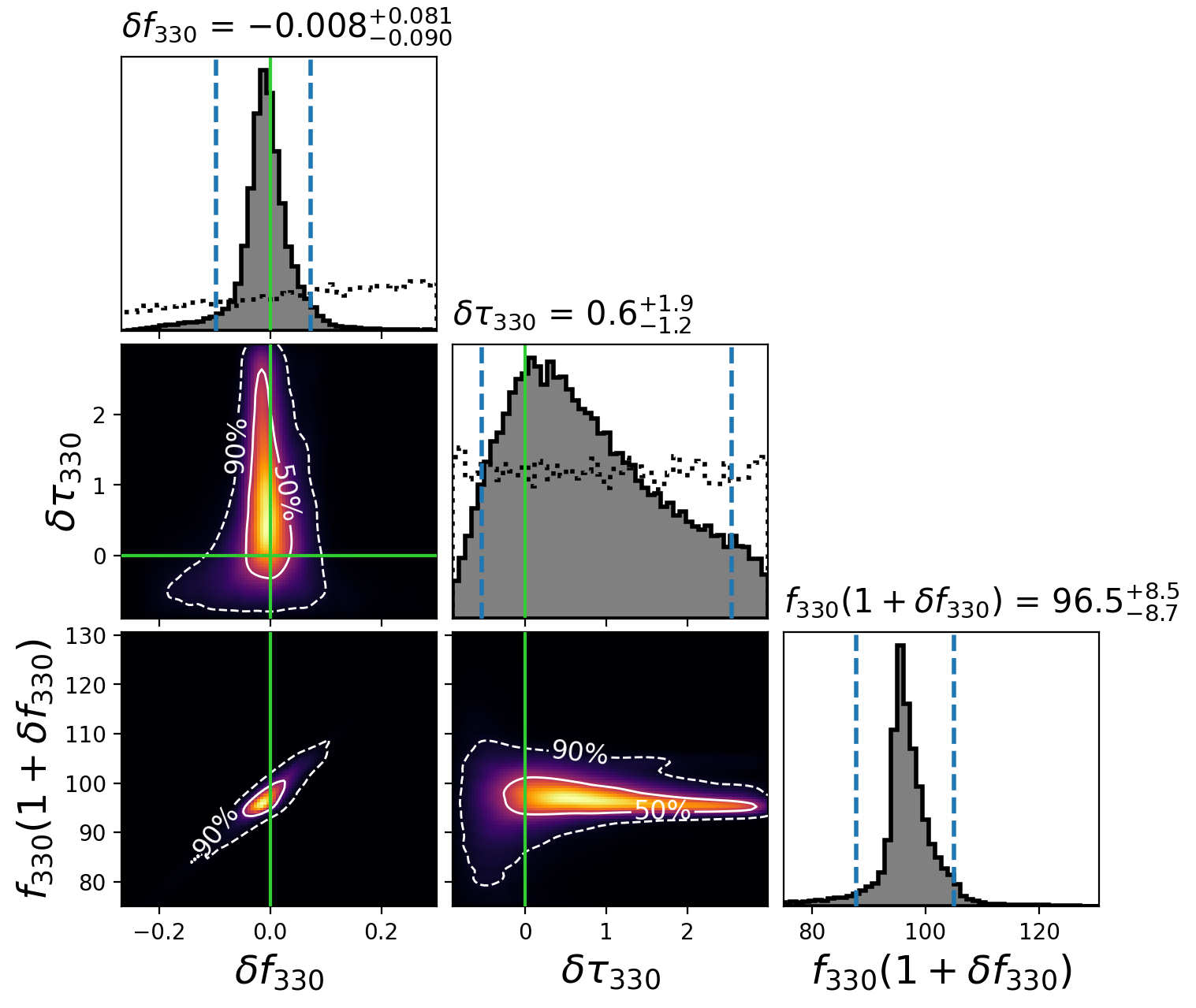}
\caption{Posterior on the deviation from Kerr of the (330) frequency $\delta f_{330}$ and damping time $\delta \tau_{330}$, as well as the resulting modified $(330)$ frequency, using a model in which we include the (220)+(330) modes at $t_{\rm ref}+6\,$ms. Quoted values are the median and 90\% credible interval, which is indicated by the dashed vertical lines. The fractional deviations are expected to be zero for a Kerr black hole (indicated by the green lines). We use a prior (black dotted lines) that is uniform over $\delta f_{330}\in[-0.3, 0.3)$, with the constraint that $f_{330}(1+\delta f_{330}) > 75$\,Hz. This constraint is necessary to avoid label switching with the (220) mode; even with the constraint we clearly measure a lower bound on $\delta f_{330}$. For the damping time we use a prior that is uniform over $\delta \tau_{330} \in [-0.9, 3)$, and find that the damping time is only weakly constrained.}
\label{supfig:delta330}
\end{figure*}

\end{document}